# Experience of Using a Multi-Cathode Counter (MCC) in a Search for Hidden Photon CDM

**A.V.Kopylov**[*], **I.V.Orekhov and V.V.Petukhov**

*Institute for Nuclear Research of Russian Academy of Sciences,*
*117312, Prospect of 60$^{th}$ October Revolution 7A, Russia*
*E-mail*: beril@inr.ru

ABSTRACT: We report on a new technique of a Multi-Cathode Counter (MCC) developed to search for hidden photon (HP) cold dark matter (CDM) with a mass from 5 to 500 eV. The method suggested in the assumption that hidden photons of the mass greater than a work function of the metal, the cathode of the counter is fabricated induce emission of single electrons from a cathode. Three configurations of the same counter are used to measure the count rates $R_1$, $R_2$ and $R_3$ of the single electron events during several weeks. As a measure of the effect from HP the difference $R_1 - (R_2 - D_3/D_2 \cdot R_3)$ has been used. Some preliminary results have been obtained on a first try to search for HP with a mass from 5 to 500 eV. The upper limit for mixing parameter χ has been set on the level of 3.3·10$^{-10}$ for hidden photon mass from 9 to 100 eV. In our plans is to refine the procedure of data treatment and to continue the measurements to collect more data.

---

[*] Corresponding author.

**Contents**

## 1. Introduction

The present observational cosmology revealed that our visible world is "immersed" in a non-baryonic matter (dark matter, DM) with mass 5 times higher than a mass of the visible world but we don't know what would be the nature of this substance. In other words we can't say what would be the matter our universe is made of. This certainly is absolutely inappropriate situation and presents probably the main challenge for experiment to-day. Theoreticians are suggesting a number of candidates for DM among them WIMPs which should produce nuclear recoils in matter topping the list. The experiments on WIMPs are now in the main stream of the search. The most exciting result has been obtained by DAMA/LIBRA collaboration. They have observed with more than 9 sigma evidence of the seasonal variation of the signal presumably originated from the vector sum of the velocities of the Earth and the Sun. This result has been obtained in 1.33 ton year of the exposure over 14 annual cycles with the sodium iodide detectors of the total mass 100 to 250 kg [1]. In spite of a very impressive statistics of this result, there's some skepticism that their signal is really from WIMPS due to growing conflicts with other experiments (CDMS, XENON10, LUX, XMASS and others) whose sensitivities now exceed the one of DAMA/LIBRA. Now the experiments of the generation 2 (G2) are on the run to cover all the natural mass range of WIMP. Among them probably the most notable is LZ project [2]. If these experiments do not observe WIMP then this may become a turning point being a serious argument that more likely a WIMP paradigm is not a real solution for DM. There are indeed other candidates, among them axion-like particles (ALP) or hidden-sector photon (HP, $X^\mu$) CDM which is light extra gauge boson. A number of experiments are searching for these objects, for example, axions are searched in the Axion Dark Matter eXperiment (ADMX) by means of a resonant cavity and magnetic field [3]. Hidden photons may be observed in experiment through a kinetic mixing $(c/2)F_{\mu\nu}X^{\mu\nu}$ with the ordinary photons. It has been shown that there is still a huge region of allowed parameters for hidden photon; see for example [4] and references therein. Recently the eV mass range of HP was investigated with a dish antenna [5], a novel method proposed in [6]. The idea is to detect the reflected from metallic mirror (antenna) electromagnetic wave which is emitted by the oscillation of electrons of the antenna's surface under the tiny electric field of the HP. This method works well only if the reflectance of mirror is high. Here in our work we focus on shorter wavelengths, i.e. higher masses of HP for which the reflectance of mirror is low. It is assumed that in this case the oscillation of electrons of the antenna's surface under the tiny electric field of HP will induce with a certain quantum efficiency η the emission of single electrons from a mirror. This value is taken to be equal to the quantum efficiency for a given metal surface and for real photons with energy $\omega = m_{\gamma'}$. Here we describe a special technique of Multi-Cathode Counter (MCC) developed for recording these events and some very preliminary data to illustrate the potential of this technique in the search for hidden photons.

## 2. Experimental apparatus

The general view of the counter is presented on Fig.1 and the electronic scheme on Fig.3. The present design of MCC first described in [7] is a further development of the work with the aim to build a device to register the neutrino- nucleus coherent scattering [8,9]. The cathode of the counter is 194 mm in diameter and 400 mm in length. It has relatively large ($\approx 0.2$ m$^2$) surface which acts in this experiment as a "mirror" for HP but instead of reflecting light it emits single electrons. The counter has a central anode wire of 20 μm and 4 cathodes, 3 of them are composed of an array of 100 μm nichrome wires tensed with a pitch of a few mm around the anode wire one after another, and a fourth one, more distant from anode, is a cathode, a "mirror", which is made of copper in the form of a cylinder.

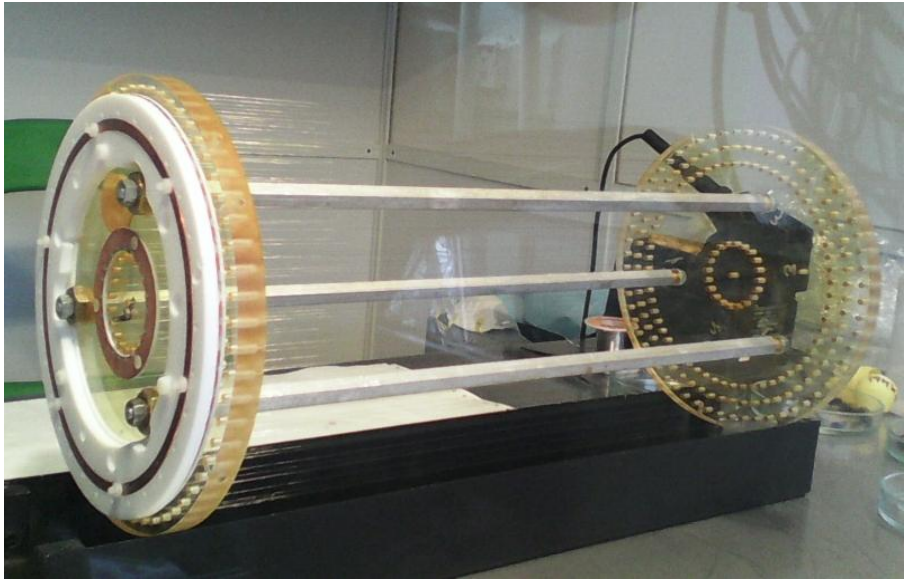

Figure 1. The central part of the counter.

The apparatus is counting electrons emitted from the walls of a "mirror" at short wavelengths $\omega = m_{\gamma'} \approx 5 - 500$ eV for which the reflectance of a "mirror" is low. The counter is filled by a gaseous mixture of argon plus 10% methane at 0.2 MPa in this particular work. The diameter of the first cathode is 40 mm to ensure high ($\approx 10^5$) coefficient of gas amplification in the central section of the counter. Second cathode of 140 mm in diameter is used for correction of the measured count rates as will be explained later. Third cathode of the diameter 180 mm is used for measurements of the background. The counter is used in three different configurations. In the first configuration 4th cathode (a copper cylinder) is under the highest negative voltage to ensure that all electrons emitted from the copper surface would drift to the central section with high gas amplification. This configuration is used to measure the count rate $R_1$ of single electrons emitted from a copper cathode. In the second configuration the highest negative voltage is applied to the 3rd cathode of diameter 180 mm. In this configuration 4th cathode, a copper cylinder, is kept at about 300 V smaller negative voltage so that electrons emitted from the copper are scattered back in argon at 0.2 MPa by high negative voltage applied to the 3rd cathode. The counter in this configuration presumably measures the background count rate $R_2$ because the geometry of the counter in this configuration is very similar to the one in the first configuration. So, opposite to a common practice of using a central part of a detecting volume as a fiducial volume, here only a small layer adjacent to the copper cathode is used to measure the effect, the rest of the sensitive volume is used for measurement of the background. It enables to subtract from the measured effect the contribution from the ends of the counter where electric fields are distorted and also the one from emission of single electrons from multiple nichrome wires. But we have additional background channel in this configuration as one can see from Fig.2.

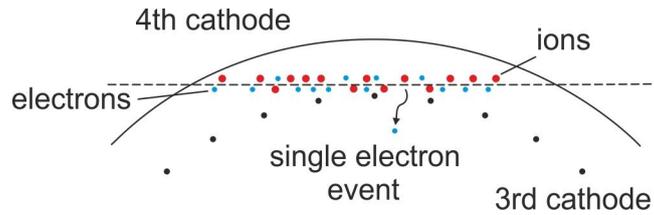

Figure 2. The background channel from tracks of ionized particles fully confined in space between 4th and 3rd cathodes.

The ionized particles with a track fully confined within a space between 4th and 3rd cathodes produce electron-ion pairs and some of these electrons diffuse into the 3rd section. To evaluate the contribution of this channel we use the 2nd cathode. Thus in the third configuration the highest negative voltage is applied to the 2nd cathode while a copper cylinder is kept under the same potential as in second configuration. The count rate measured in this configuration should also have the background channel depicted on Fig.2 but lower than the one in 2nd configuration by a factor of $D_2/D_3$ where $D_2$ and $D_3$ are diameters of the second and third cathodes. So as a measure of the effect from single electrons emitted from a surface of copper cathode we used the expression: $R_{MCC} = R_1 - (R_2 - D_3/D_2 \cdot R_3)$.

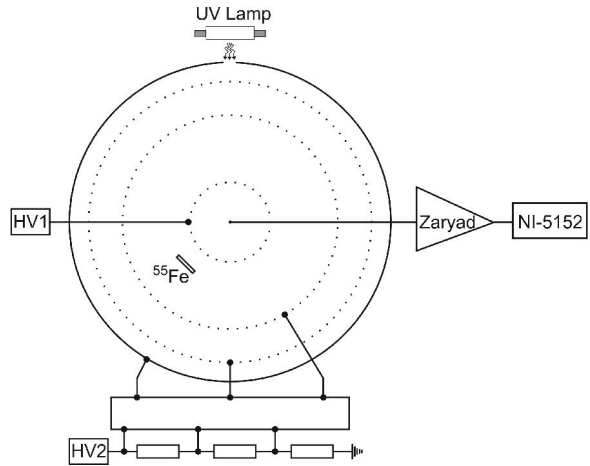

Figure 3. The simplified electronic scheme of a multicathode counter (MCC).

The measurements of $R_1$, $R_2$ and $R_3$ are performed sequentially one after another. After refill of the counter it took about 3 weeks of work under high voltage before counter has got stabilized. Presumably this happens as a result of some purification of a copper cathode by electric discharges initiated by ionized particles crossing the counter. In the measurements the shapes of the pulses on the output of a charge sensitive preamplifier are recorded by 8-bit digitizer. Pulse shape discrimination is used to select "true" pulses from "noise" pulses similar to what has been described in [8, 9]. The "true" pulses have typically a relatively short front edge (a few microseconds) corresponding to the drift of positive ions to cathode and long (hundreds of microseconds) tail corresponding to the time of the baseline restoration of the charge sensitive preamplifier. The "noise" pulses usually had a wrong (too fast or too slow or irregular) front edge. In the analyses of the data only the pulses with a baseline within ± 2 mV were taken into analyses with a proper evaluation of the resulted live time. To reduce the background from external γ-radiation the counter has been placed in a steel cabinet with 30 cm iron shield. It enabled to decrease the count rates of single-electron events by a factor of 2 (detector on a porch of a steel cabinet versus detector inside a steel cabinet) while the flux of gamma rays in the region between 511 and 661.6 keV lines was attenuated in these lay-outs by a factor of 30.

## 3. Energy calibration and analysis

The calibration of the counter has been conducted by $^{55}$Fe source and by UV light of the mercury lamp. The source has been placed inside the counter between first and second cathodes facing the anode wire. High voltage at first cathode was 2060 V and high voltages applied to voltage divider has been used for all three configurations such as to ensure the amplitude of the pulse corresponding to 5.9 keV of $^{55}$Fe on the output of charge sensitive preamplifier to be at the level 1400 mV what corresponds to a gas amplification of about $10^5$. The obtained spectrum from $^{55}$Fe source is presented at Fig.4. One can see the nonlinearity in the spectrum which indicates the regime of limited proportionality. Figure 5 shows the approximation voltage versus energy by two peaks: 5.9 keV and 2.95 keV. From this approximation one can see that at energies less than 100 eV the conversion factor is ≈ 2.27 eV/mV.

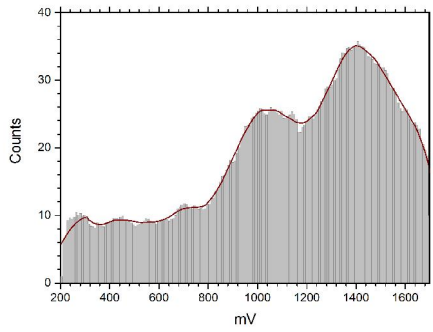
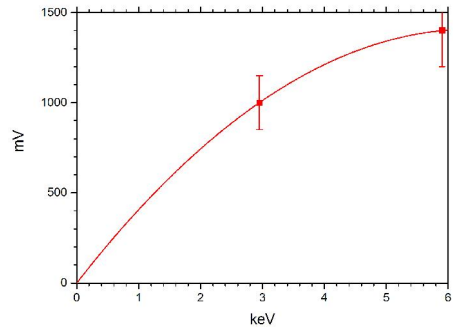

Figure 4. The spectrum from $^{55}$Fe source. peaks.

Figure 5. The approximation V/E by two

It means that the conversion factor in counting the single electron events by our detector can be taken equal to 2.27 eV/mV. Then the source has been removed and the internal walls of the counter were irradiated by UV light from a mercury lamp placed outside through a window made of melted silica. Figure 6 shows the single electron spectra obtained in measurements in 1$^{st}$ and 2$^{nd}$ configurations.

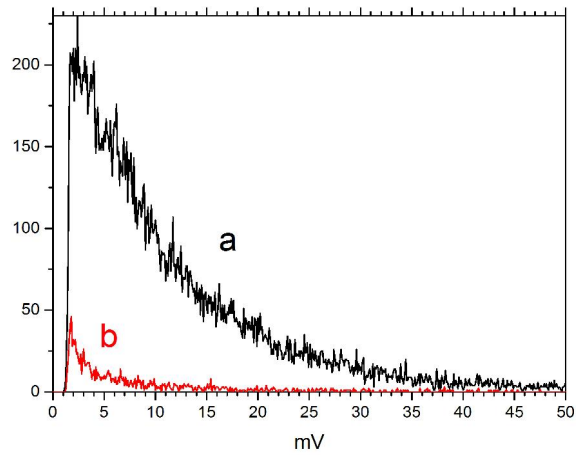

Figure 6. The single electron spectra obtained in measurements in 1$^{st}$ (a) and 2$^{nd}$ (b) configurations at the same flux of UV photons. The conversion factor is ≈ 2.27 eV/mV.

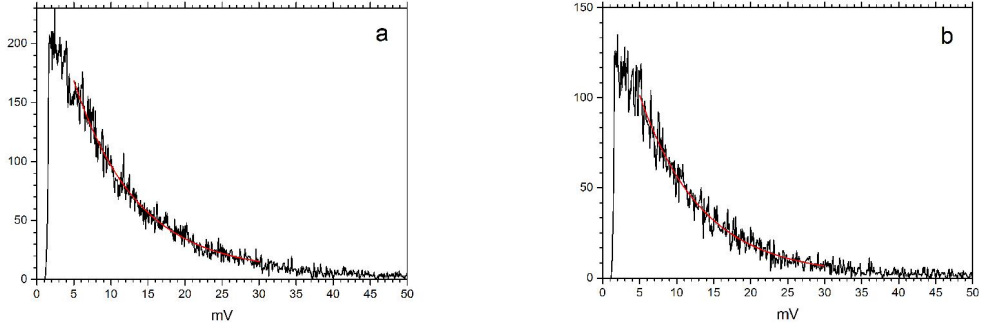

Figure 7. The single electron spectra obtained in measurements in 1st (a) and 2nd (b) configurations at the similar statistics collected. The conversion factor is ≈ 2.27 eV/mV.

Comparing the count rates $R_1$ and $R_2$ presented on Fig.6 one can see that in the 2nd configuration the count rate of single electrons was about 10 times lower than in the first one. It proves that in the 2nd configuration the electrons emitted from the walls of the counter were really rejected back by the 3rd cathode. It means that the counter in the 2nd configuration can be used for measurement of the background of the single electron count rates measured in the first configuration. Figures 7 shows the single electron spectra obtained in 1st and 2nd configurations by approximately the same statistics. One can see that both spectra have exponential shape and inverse indexes of exponents are equal within the uncertainties of measurements: 8.4 ± 0.4 mV and 8.5 ± 0.3 mV. Taking into consideration the conversion factor 2.27 eV/mV one can see that they are very close to the average energy 27 eV to produce electron-ion pare in argon. The data were collected frame by frame. Each frame contained 2M points each point 100 ns. After collection of the data they were stored on a disk then the collection resumed. The analysis of the collected data was performed off-line. It consists in searching for small peaks of the "true" shape with the amplitude from 3 to 30 mV which corresponds to energies from 6.8 to 68 eV. Only pulses with a baseline within ± 2 mV were taken into analyses with proper evaluation of a live time in each measurement. The frames with the signs of excessive noisiness were removed from analysis.

## 4. Sensitivity of the method

The experiment was conducted in the assumption that hidden photons of the mass greater than a work function of a metal, the cathode of the counter is fabricated induce the oscillation of the current on the surface of a cathode with the resulted emission of electrons from the surface. The substantial point here is that the conversion of HP into free electron occurs at the interface of metal – dielectric. The probability for this oscillation of the current induced by hidden photon to emit electron was taken to be equal to quantum efficiency η for the UV photon. If we take this then we can easily transform the sensitivity from the case of a dish antenna to our case of the counter with a cathode as a "mirror". Taking the sensitivity from [6], this is written here for the case when all dark matter is composed of hidden photons,

$$\chi_{sens} = 5.6 \times 10^{-12} \left(\frac{R_{\gamma,det}}{1\ \mathrm{Hz}}\right)^{1/2} \left(\frac{m_{\gamma'}}{1\ \mathrm{eV}}\right)^{1/2} \left(\frac{0.3\ GeV/cm^3}{\rho_{CDM,halo}}\right)^{1/2} \left(\frac{1\ \mathrm{m}^2}{A_{dish}}\right)^{1/2} \left(\frac{\sqrt{2/3}}{\alpha}\right)^{1/2} \quad (1)$$

where $R_{\gamma,det}$ is the minimum count rate from HP which can be observed in the experiment. One can see that for $m_{\gamma'}$ =10 eV and $A_{dish}$ = 0.2 m$^2$ to set a limit on the level of $10^{-11}$ one should have $R_{\gamma,det}$ on the level of 0.1 Hz. The value $R_{MCC}$ should be in this case lower than this value by a factor $1/\eta$. In case of a dish antenna η is a quantum efficiency of a photocathode of PMT and according to [5] PMT they used had a sensitive range 300-650 nm and a peak quantum efficiency of 17%. In our case at the energies $E_\gamma = m_{\gamma'} \approx$ 10 eV which correspond to wavelength 120 nm the probability that electron will be ejected η ≈ 1%. Thus the value $R_{MCC}$ should be in this case on the level of $10^{-3}$ Hz. The sensitivity depends on η as a square root but certainly, in comparison with a dish antenna we somewhat loose in sensitivity due to lower quantum efficiency η. We propose here a complimentary method: in comparison with a dish antenna we can cover adjacent broad region of higher masses for HP and if we use for the cathode of the counter several materials with different Z it will be possible to look into narrow intervals of the mass range from 5 to 500 eV.

## 5. First data obtained

The measurements were performed during 52 days after the counter got refilled. It took three weeks before the counter has got stabilized. We explain it by purification of a copper cathode in electric discharges. The count rate *R2* is regularly higher than $R_1$. This partially can be explained by the additional background channel depicted on Fig.2. All count rates are in a few Hz range. The scattering of the points determine the total uncertainties of measurements. These uncertainties were used in finding uncertainty of the average value for each rate $R_1$, $R_2$ and $R_3$. The average value of $R_{MCC}$ calculated for 31 days of stable work of the counter was found to be: $\overline{R}_{MCC}$ = - 0.33 ± 0.41 Hz. The uncertainty has been found from the real scattering of the experimental points. So if we take the normal distribution for uncertainties, then we obtain that at 95% confidence level: $\lim \overline{R}_{MCC}$ < 0.49 Hz. The probability for the oscillation of the current induced by HPs to emit an electron from the surface of a metal was taken to be equal to quantum efficiency η taken from [10]. Using the expression (1) we obtained a limit for a mixing constant χ. Here: $R_{\gamma,det} = \lim \overline{R}_{MCC} /\eta$. Figure 8 shows the obtained limit for χ as a function of the mass of the hidden photon.

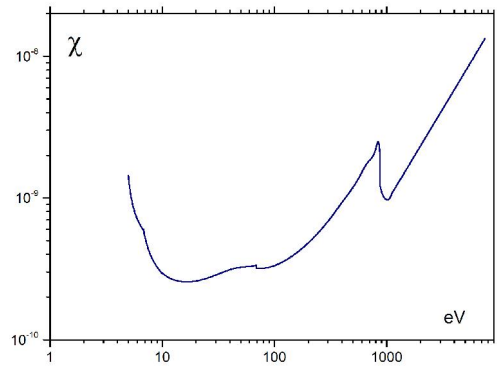

Figure 8. The limit for a mixing constant χ.

These numbers are first, very preliminary result for hidden photons with a mass from 5 to 500 eV. Stellar astrophysics provides stringent constraints for this value. Our result is deep inside the regions excluded by astrophysics, see, for example, [11] and references therein. But this result has been obtained in a direct measurement experiment by observing single electrons emitted from the surface of a metal cathode and because of this we believe that it deserves attention. We plan to continue the measurements to collect more data and in our plans also is to perform the measurements with the cathodes made of metals with different Z to look whether the effect depends on the metal the cathode of the counter is made of.

## 6. Conclusion

A new technique of Multi Cathode Counter (MCC) has been developed to search for hidden photon CDM in the assumption that all dark matter is composed of hidden photons (HP). It was assumed also that HPs of the mass greater than a work function of the metal, the cathode of the counter is fabricated induce emission of single electrons from a cathode. First preliminary result has been obtained for HPs with a mass from 5 to 500 eV. In our plans are to continue the measurements to collect more data and to refine the procedure of data treatment. At present time we also construct a new detector with a more developed design.

## Acknowledgements

The authors express deep gratitude to E.P.Petrov for continuous help during this work, to A.I.Egorov for skill and patience during fabrication of many different versions of the counter and to Grant of Russian Government "Leading Scientific Schools of Russia" #3110.2014.2 for partial support of this work.